\documentclass[10pt, twocolumn, floatfix, superscriptaddress, aps, prb]{revtex4-2}
\usepackage{amsmath, amssymb, graphicx, hyperref, xcolor, colortbl, braket, bbold}
\newcommand{\changes}[1]{#1}

\begin{document}

\title{Unscrambling of single-particle wave functions \texorpdfstring{\\}{}in systems localized through disorder and monitoring}

\author{Marcin Szyniszewski}
\affiliation{Department of Physics and Astronomy, University College London, Gower Street, London, WC1E 6BT, United Kingdom}

\begin{abstract}
  In systems undergoing localization-delocalization quantum phase transitions due to disorder or monitoring, there is a crucial need for robust methods capable of distinguishing phases and uncovering their intrinsic properties. In this work, we develop a process of finding a Slater determinant representation of free-fermion wave functions that accurately characterizes localized particles, a procedure we dub ``unscrambling''. The central idea is to minimize the overlap between envelopes of single-particle wave functions or, equivalently, to maximize the inverse participation ratio of each orbital. This numerically efficient methodology can differentiate between distinct types of wave functions: exponentially localized, power-law localized, and conformal critical, also revealing the underlying physics of these states. \changes{The method is readily extendable to systems in higher dimensions.} Furthermore, we apply this approach to a more challenging problem involving disordered monitored free fermions in one dimension, where the unscrambling process unveils the presence of a conformal critical phase and a localized area-law quantum Zeno phase. Importantly, our method can also be extended to free fermion systems without particle number conservation, which we demonstrate by estimating the phase diagram of $\mathbb{Z}_2$-symmetric disordered monitored free fermions. Our results unlock the potential of utilizing single-particle wave functions to gain valuable insights into the localization transition properties in systems such as monitored free fermions and disordered models.
\end{abstract}

{\maketitle}

\section{Introduction}

In closed quantum mechanical systems, unitary evolution often leads to thermalization, a process where information about initial conditions becomes inaccessible in local observables~\cite{Deutsch1991, Srednicki1994, DAlessio2016, Borgonovi2016}. There are however intriguing exceptions, where the system instead exhibits non-equilibrium behavior, such as retaining useful information for arbitrarily long times. Paradigmatic examples of this are localized systems, where the disorder is responsible for the breaking of the eigenstate thermalization hypothesis.

Within the context of many-body systems, the transition between a many-body localized (MBL) phase and a thermalized phase has captivated researchers and sparked debates~\cite{Basko2006, Gornyi2005, Pal2010, Nandkishore2015, Abanin2019}. The prospect of stabilizing quantum information using disorder is particularly relevant for experiments involving quantum memories, and quantum computation in general. From the \changes{conceptual} point of view, the transition to the localized phase can be characterized by properties of the energy spectrum and quantum states. In the former, one often compares the system to the relevant random matrix theory ensemble, examining statistics such as nearby level spacings~\cite{Haake2010, Oganesyan2007, Atas2013}. On the other hand, assessing the localization properties of wave functions can be done through inverse participation ratio (IPR), a powerful tool in the study of MBL systems~\cite{DeLuca2013, Luitz2015, Calixto2015, Misguich2016, Bera2015}.

In the case of free fermion systems, IPR has proven effective in characterizing the Anderson localization transition~\cite{Kramer1993, Wegner1980, Bauer1990, Fyodorov1992}, where the single-particle wave functions \changes{become exponentially localized in the insulating phase. Other non-equilibrium processes in free fermion systems have seen a resurgence in interest, owing largely to their tractability through the single-particle picture. Most prominent examples include dynamical phase transitions~\cite{Heyl2013, Heyl2015, Zvyagin2016, Heyl2018}, time crystals~\cite{Khemani2016, Russomanno2020}, quantum quenches~\cite{Rossini2009, Rossini2010, Calabrese2011, Calabrese2012pt1, Calabrese2012pt2, Ziraldo2012, Ziraldo2013}, Floquet engineering~\cite{Russomanno2012, Bhattacharyya2012, Russomanno2013, Lazarides2014, Sharma2014, Russomanno2017}, and monitored quantum systems~\cite{Cao2019, Alberton2021}. The last have garnered attention in the context of measurement-induced entanglement transitions~\cite{Nahum2017, Skinner2019, Chan2019, Li2018, Szyniszewski2019, Li2019}, where repeated measurements force the system into an area-law quantum Zeno phase. Notably, albeit} the single-particle wave functions serve as a valuable theoretical tool for visualizing and understanding free-fermion physics, their precise definition is somewhat ambiguous, which leaves the physics unchanged but may hamper the theoretical interpretation. As we will see in the course of this paper, we can resolve this ambiguity by establishing a clear link between the procedure for obtaining relevant single-particle representations and the IPR.

\begin{figure}[b]
  \includegraphics[width=1\columnwidth]{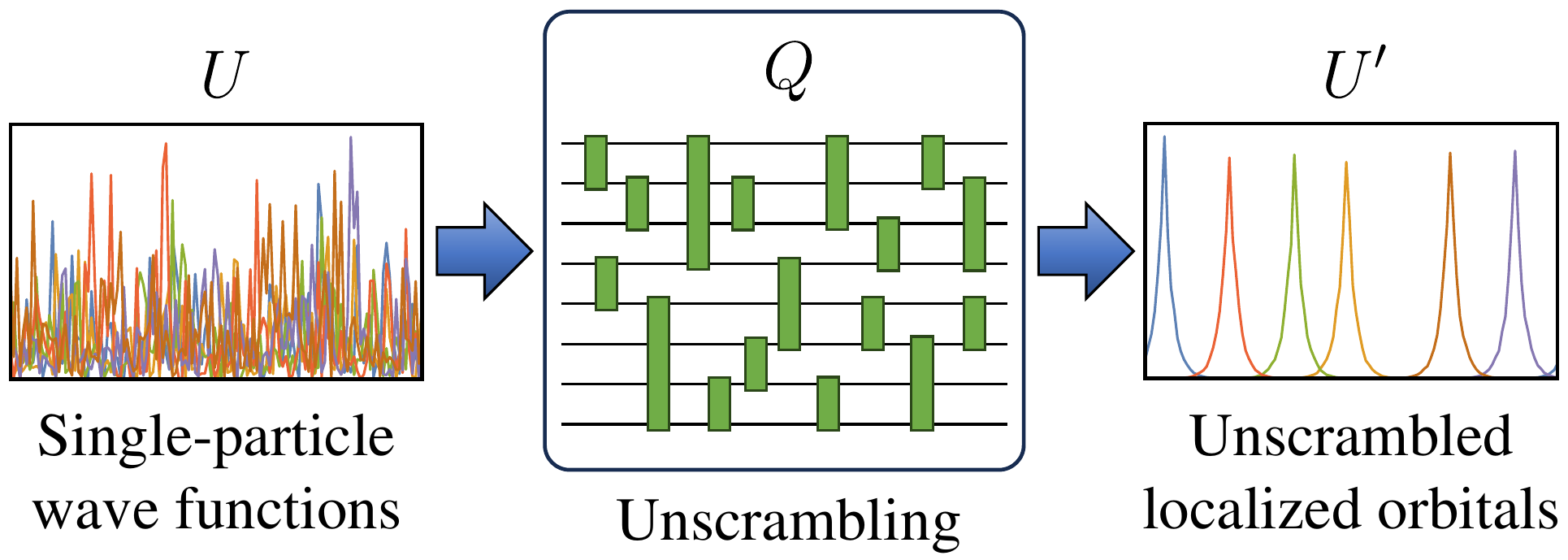}
  \caption{A diagrammatic picture of the unscrambling method $Q$, illustrating how the original Slater determinant matrix $U$ transforms into a collection of localized orbitals $U' = U Q$. \changes{$U$ and $U'$ are represented by plotting the density of the corresponding orbitals as a function of a position. $Q$ is approximated as a series of two-orbital transformations (green gates).}\label{fig:summary}}
\end{figure}

Consider a free-fermion wave function $| \psi \rangle$, subject to particle number conservation. This wave function can be fully characterized by its correlation matrix $D$, with elements $D_{i j} = \langle c_i^{\dag} c_j \rangle$. Alternatively, we can express the wave function using an ansatz,
\begin{equation}
  | \psi \rangle = \prod_{n = 1}^N \left( \sum_{i = 1}^L U_{i n} c_i^{\dag} \right) | 0 \rangle,
  \label{eq:u1ansatz}
\end{equation}
where $N$ represents the number of particles, $L$ denotes the system size, and $| 0 \rangle$ is the vacuum state. It is easy to see that matrix $U$ can be interpreted as a Slater determinant, where each column of $U$ corresponds to a single-particle wave function $|\psi_n\rangle$ (also known as an orbital). Notably, matrix $U$ is an isometry, $U^{\dag} U = \mathbb{1}$, and is intrinsically linked to the correlation matrix, as $D = U U^{\dag}$. However, it is essential to notice that the matrix $U$ is not uniquely defined, as a multiplication $U' = U Q$ by any unitary matrix $Q$, leaves the correlation matrix (and thus the physical state itself) unchanged.

Hence, we pose the central problem of this study: is there a method of transforming the Slater determinant matrix $U$ into a set of single-particle wave functions that are endowed with specific properties, such as localization? To answer this question, we propose a methodology based on the average IPR of all \changes{particle orbitals} in the system. This procedure aims to yield single-particle wave functions that reveal insights into the presence or absence of localization (as illustrated in Fig.~\ref{fig:summary}), as well as internal model-specific characteristics. Furthermore, we intend to validate this method by applying it to wave functions produced through the interplay of disorder and monitoring\changes{, where a measurement-induced entanglement transition is present}. Our objective is to demonstrate that the proposed approach is applicable to a diverse range of systems, extending beyond those solely localized through the disorder.

The paper is structured as follows. In Sec.~\ref{sec:methodology} we propose a methodology that approximates matrix $Q$ needed to transform the representation into localized single-particle orbitals. We test this method in Sec.~\ref{sec:testing} on three models exhibiting three distinct behaviors: exponentially localized, conformally invariant, and power-law localized. \changes{We apply the method to a higher-dimensional system in Sec.~\ref{sec:3d-anderson}, showing its versatility across different lattice geometries.} In Sec.~\ref{sec:monitored} we show how the method behaves in a monitored disordered system with a measurement-induced phase transition between localized and critical regimes. We also extend the applicability of this methodology to systems without number conservation, as discussed in Sec.~\ref{sec:z2}. Finally, we conclude in Sec.~\ref{sec:discussion}.

\section{Methodology}
\label{sec:methodology}

We begin by addressing the two-particle problem and subsequently extend this approach to a multi-particle solution. Our ultimate goal is to find the inverse of matrix $Q^{-1} = Q^\dagger$, which can be thought of as a transformation that scrambles useful information contained in localized orbitals. Hence, we aptly term the process of finding $Q$ ``unscrambling''.

\subsection{The two-particle problem}

When considering only two particles, a possible transformation $Q$ takes the form of a generic $2 \times 2$ unitary matrix,
\begin{equation}
  Q = e^{i \alpha / 2} \left(\begin{array}{cc}
    e^{i \varphi} \cos \theta & e^{i \beta} \sin \theta\\
    - e^{- i \beta} \sin \theta & e^{- i \varphi} \cos \theta
  \end{array}\right),
\end{equation}
which depends on four free parameters $(\alpha, \beta, \varphi, \theta)$. However, one can notice that any overall phase of a single-particle wave function does not alter its physical interpretation. Consequently, we can reduce the number of free parameters to just two:
\begin{equation}
  Q (\varphi, \theta) = \left(\begin{array}{cc}
    e^{i \varphi} \cos \theta & \sin \theta\\
    - \sin \theta & e^{- i \varphi} \cos \theta
  \end{array}\right) .
\end{equation}

\begin{figure}[tb]
  \includegraphics[width=1\columnwidth]{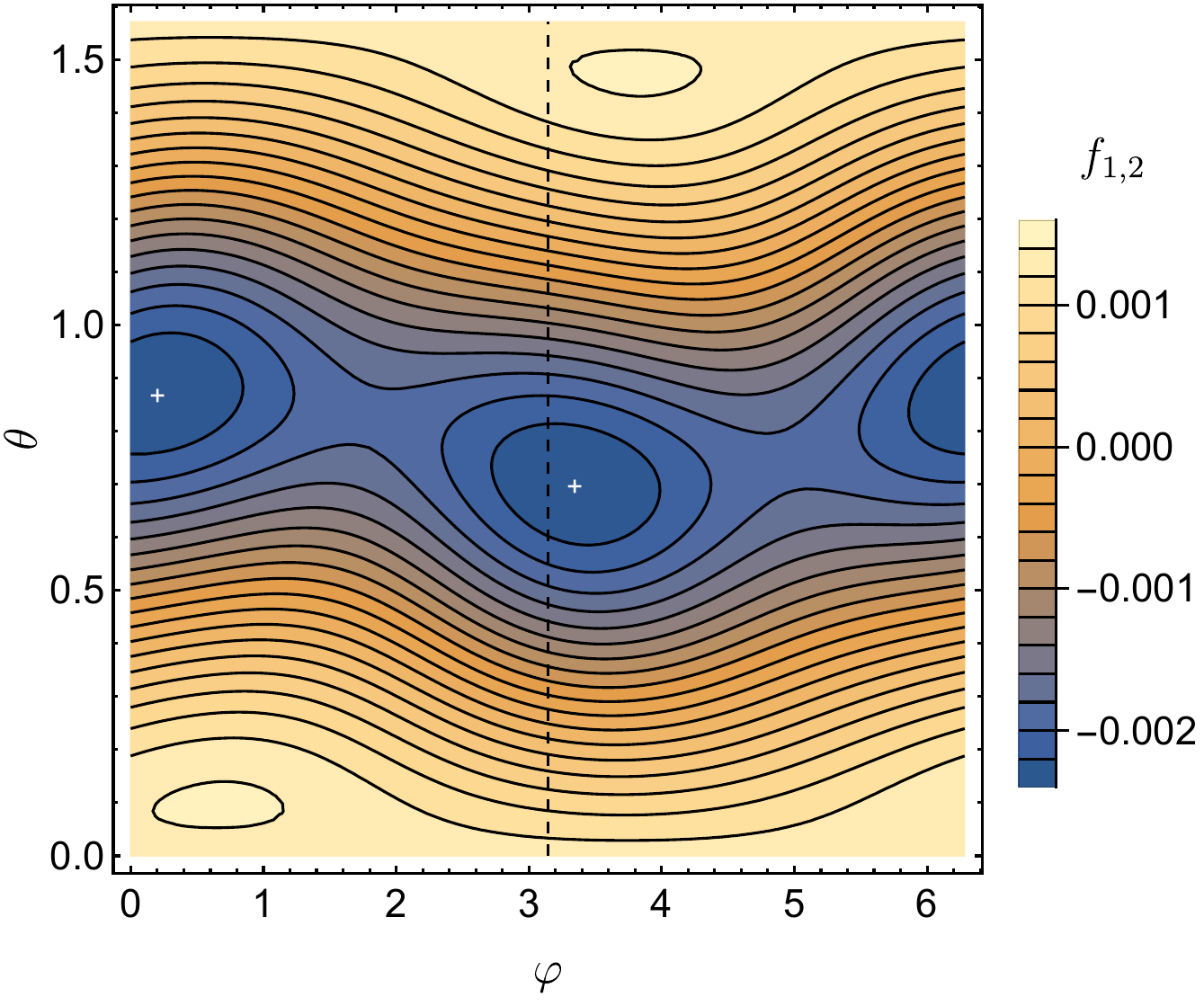}
  \caption{Example of the cost function $f_{1, 2} (\varphi, \theta)$ showing periodicity in both variables. The dashed line divides two parts which are reflections of each other. The minima are designated by white crosses.\label{fig:cost}}
\end{figure}

Now, we introduce a cost function that encompasses the localization properties of the two particles. We propose minimizing the overlap between the envelopes of single-particle wave functions. For two particles labeled 1 and 2, this cost function is defined as
\begin{equation}
  f_{1, 2} (\varphi, \theta) = \sum_i | U_{i 1}' |^2 | U_{i 2}' |^2 .
\end{equation}
The computational advantage lies in rearranging the expression by pushing the sum inside, yielding
\begin{align}
  f_{1, 2} (\varphi, \theta) & = A \cos^2 (2 \theta) + \text{Re} [B e^{i \varphi}] \sin (2 \theta) \cos (2 \theta) \\
  & \quad - \text{Re} [C e^{2 i \varphi}] \sin^2 (2 \theta) + D \sin^2 (2 \theta), \nonumber
\end{align}
with
\begin{align}
  A & = \sum_i | U_{i 1} |^2 | U_{i 2} |^2, \\
  B & = \sum_i (| U_{i 1} |^2 - | U_{i 2} |^2) U_{i 1} U_{i 2}^{\ast}, \\
  C & = \frac{1}{2} \sum_i U_{i 1}^2 (U_{i 2}^{\ast})^2, \\
  D & = \frac{1}{4} \sum_i (| U_{i 1} |^4 + | U_{i 2} |^4) .
\end{align}
Coefficients $A, B, C, D$ can be calculated once before the minimization process. Notably, the cost function $f_{1, 2} (\varphi, \theta)$ (see Fig.~\ref{fig:cost} for an example) is periodic in both $\varphi$ and $\theta$, with periods $2 \pi$ and $\pi / 2$ respectively, and has one minimum in the region $\varphi \in [0, \pi), \theta \in [0, \pi / 2)$. As a result, the computational effort of finding the minimum of $f_{1, 2}$ numerically is relatively low.

\subsection{The multi-particle problem}

When dealing with a large number of particles ($N$), a challenge arises in parametrizing a generic unitary matrix $Q$. The complexity stems from the fact that the number of free parameters grows as $(N^2 - N)$. Here, the subtracted $N$ accounts for the inherent freedom in choosing single-particle wave function phase factors. To circumvent this computational hurdle, we propose to approximate $Q$ as a series of two-particle transformations.

To achieve our goal, we minimize each pair of particles consecutively, until a convergence is reached for the following cost function,
\begin{equation}
  f = \sum_{n < m} f_{n, m},
  \label{eq:cost}
\end{equation}
which is simply a sum of all two-particle cost functions. Crucially, this cost function never increases during any two-particle minimization. To understand why, let us examine the change in $f$ when minimizing particles $n$ and $m$:
\begin{align}
  \delta f & = f' - f \\
  & = \delta f_{n, m} + \sum_{n' \neq \{ n, m \}} \delta f_{n, n'} + \sum_{n' \neq \{ n, m \}} \delta f_{m, n'} \\
  & = \delta f_{n, m} \\
  & \quad + \sum_i (| U_{i n}' |^2 + | U_{i m}' |^2 - | U_{i n} |^2 - | U_{i m} |^2) \sum_{n'} | U_{i n'} |^2 \nonumber\\
  & = \delta f_{n, m},
\end{align}
as the expression in the bracket is zero. Hence, the proposed process is not only numerically efficient, but also ensures that the cost function never increases, with the change $\delta f$ being a local property. We also find numerically that the results of the unscrambling always correspond to local minima of the multi-particle cost function (see Appendix~\ref{app:local_min}).

We now show the intimate connection between the proposed cost function and IPR. Note that the following expression is constant during the minimization process,
\begin{align}
  \sum_i D_{i i}^2 & = \text{const.} \\
  & = 2 \sum_{n < m} f_{n, m} + \sum_n \sum_i | U_{i n} |^4 \\
  & = 2 f + \sum_n \text{IPR} (n), 
\end{align}
where IPR of a \changes{single-particle wave function} is defined as $\text{IPR} (n) = \sum_i | U_{i n} |^4 = \sum_i | \langle i | \psi_n \rangle |^4$. IPR serves as a measure of localization~\cite{Kramer1993, Wegner1980, Bauer1990, Fyodorov1992}, and can take values between 0 and 1. When IPR is large, the orbital is localized in the given basis (in the real space, $\text{IPR} \sim 1/\xi\sim \text{const.}$ for localization length $\xi$), while if IPR is approximately $1/L$, the orbital is delocalized \changes{(which is a necessary, but not a sufficient condition for thermalization)}. This gives the physical interpretation of the unscrambling method: by minimizing the global cost function $f$, which represents sums of overlaps between single-particle wave function envelopes, we maximize the average IPR [defined as $\sum_n \text{IPR} (n)/N$] of the particle orbitals. Intuitively, this approach ensures that the orbitals become as localized as possible.

\section{Testing the method}
\label{sec:testing}

In this section, we demonstrate that the introduced methodology is capable of distinguishing among various types of wave functions: extended, exponentially localized, and power-law localized. Furthermore, this approach provides insights into the internal structures of these distinct wave function behaviors.

\subsection{Exponentially localized wave functions in Anderson localized systems}
\label{sec:anderson}

We commence by applying the unscrambling procedure to the wave functions of an Anderson-localized model on a periodic chain. The model Hamiltonian takes the form
\begin{equation}
  H = \sum_{i = 1}^L \left( c_i^{\dag} c_{i + 1} + \text{h.c.} \right) + \sum_{i = 1}^L h_i n_i,
  \label{eq:anderson}
\end{equation}
where $h_i \in [- W, W]$ is a random disordered potential with a uniform box distribution and disorder strength $W$. We initialize the system in the N\'eel state with $N = L / 2$ particles, which corresponds to $U_{i n} = \delta_{i, 2 n}$. Next, we evolve the state in time, using
\begin{equation}
  U (t + d t) = e^{-i\,d t\,\mathbb{H}} U (t), \label{eq:tev}
\end{equation}
where $\mathbb{H}$ is an $L \times L$ matrix representing the Hamiltonian, with elements $\mathbb{H}_{i j} = \delta_{i, j + 1} + \delta_{i, j - 1} + h_i \delta_{i j}$. We set the time step to $d t = 0.05$. At long times, this evolution should lead to a wave function, which can be represented as a Slater determinant of exponentially localized single-particle wave functions. We, therefore, evolve the system up to $t_{\text{fin}} = 100$ and subsequently apply the unscrambling method. Note that alternatively, we could use the correlation matrix $D$ and evolve it as $D(t+dt) = \exp(-i\,d t\,\mathbb{H}) D \exp(i\,d t\,\mathbb{H})$, then subsequently use singular value decomposition to obtain $U$. However, applying operations directly on $U$ is numerically much more efficient.

\begin{figure}[tb]
  \includegraphics[width=1\columnwidth]{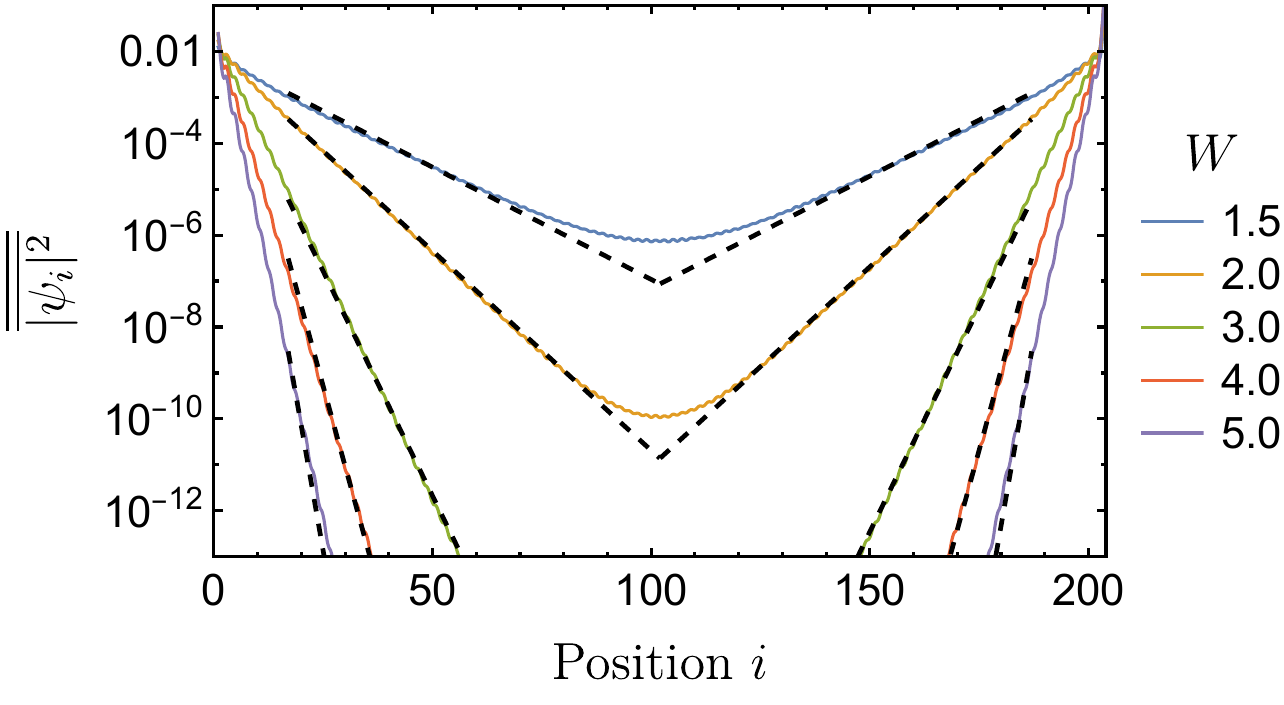}
  \caption{Anderson localization of single-particle wave functions revealed after the unscrambling procedure. The dashed lines is the expected behavior of $\sim \exp (- | i | / \xi)$, with \changes{$\xi \approx 20 / W^2$}. $L = 204, d t = 0.05, t_{\text{fin}} = 100$.\label{fig:anderson} }
\end{figure}

The resulting single-particle wave functions for an example system size of $L = 204$ are plotted in Fig.~\ref{fig:anderson}. Each wave function is centered so that the site with the largest occupation probability is at $i = 0 \equiv L$. Then we define the typical average,
\begin{equation}
  \overline{\overline{| \psi_i |^2}} = \exp \left[\overline{\ln (| U_{i n} |^2)}\right],
\end{equation}
where $\bar{\ast}$ is an average over 100 disorder realizations and all particle \changes{orbitals} $n$. \changes{This method of averaging (i.e.\@ centering first, then taking the typical average) will be used throughout the rest of this paper.}

The resulting orbitals exhibit the expected exponential localization, $\overline{\overline{| \psi_i |^2}} \sim \exp (- | i | / \xi)$. \changes{The behavior of the localization length $\xi \approx 20 / W^2$ is in complete agreement with the well-known behavior (see the dashed lines in the figure) for Anderson localization, including the prefactor~\cite{Kappus1981, Kramer1993}.} This simple test shows that the unscrambling procedure works well for exponentially-localized wave functions, not only revealing localization, but also correctly predicting the value of the localization length.

\subsection{Conformally-invariant wave functions in the XX model}
\label{sec:xx}

To test how the method deals with critical wave functions, we choose to investigate the ground state of the XX model with a transverse field, which in the fermionic language can be described by the Hamiltonian~\cite{Sachdev2011, Christe1994, Suzuki2013, Latorre2009},
\begin{equation}
  H = - \sum_{i = 1}^L \left( c_i^{\dag} c_{i + 1} + \text{h.c.} \right) + h \sum_{i = 1}^L n_i.
\end{equation}
This Hamiltonian can be easily diagonalized through the Jordan-Wigner transformation,
\begin{equation}
  H = \sum_k \Lambda_k c_k^{\dag} c_k,
\end{equation}
where the energy associated with mode $k \in \{ 0, \ldots, L - 1 \}$ is
\begin{equation}
  \Lambda_k = 2 \cos \left( \frac{2 \pi k}{L} \right) + h.
\end{equation}
The ground state can be described as the state where all occupied modes have a negative energy. Let us designate the collection of occupied modes as $\mathcal{K} \in [k_s, L - k_s]$, where
\begin{equation}
  k_s = \frac{L}{2 \pi} \arccos \left( - \frac{h}{2} \right) .
\end{equation}
It is easy to see that the two-point correlation function of the ground state is then described by
\begin{equation}
  D_{i j} = \frac{1}{L} \sum_{k \in \mathcal{K}} \cos \left( \frac{2 \pi}{L} k (i - j) \right) . \label{eq:corrXX}
\end{equation}

\begin{figure}[tb]
  \includegraphics[width=1\columnwidth]{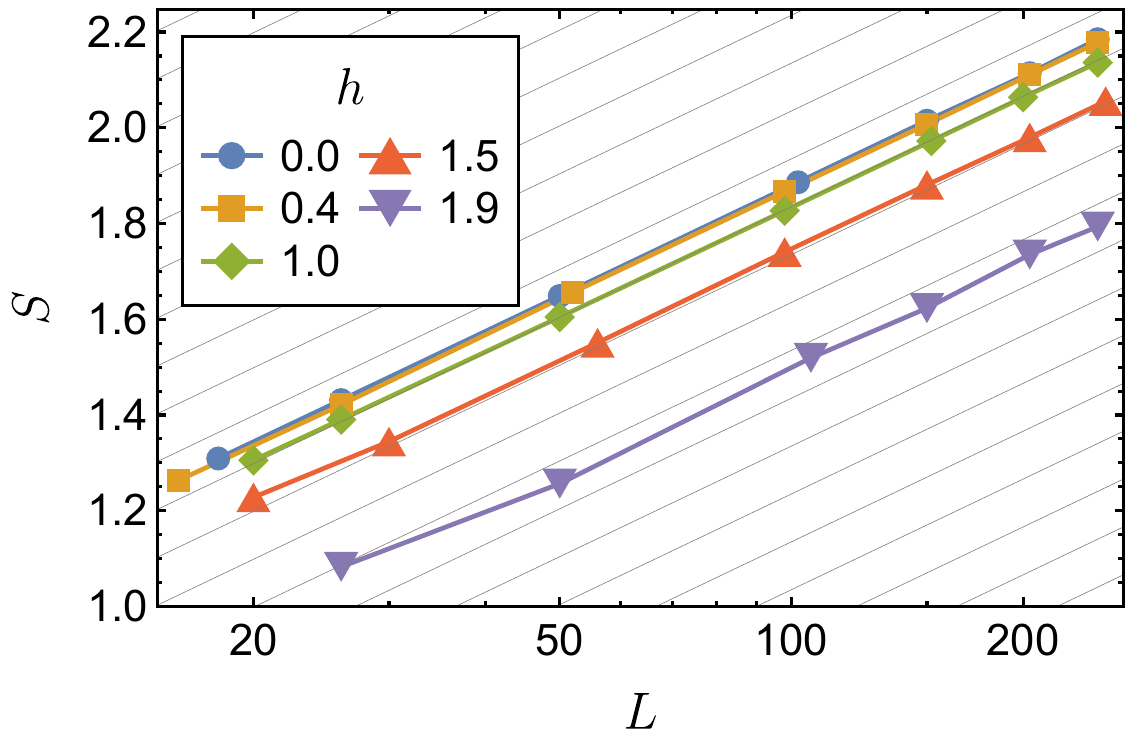}
  \caption{Entanglement entropy $S$ of the ground state of the XX model as a function of the system size $L$ and for different values of the transverse field $h$. The gray lines indicate $[(1 / 3) \ln L + C]$ behavior for different offsets $C$, signifying conformal symmetry with central charge $c = 1$.\label{fig:XXentropy}}
\end{figure}

The ground state of the XX model is conformally invariant when $| h | < 2$. We can use the correlation matrix $D$ to obtain entanglement entropy for the ground state as a function of the system size and check for conformal symmetry. The von Neumann entropy $S$ of region A can be calculated as
\begin{equation}
  S = \sum_i \left[- \lambda_i \ln \lambda_i \vphantom{\frac{}{}}- (1 - \lambda_i) \ln (1 - \lambda_i)\right],
\end{equation}
where $\lambda_i$ are eigenvalues of the correlation matrix $D$ with indices restricted to region A.

Fig.~\ref{fig:XXentropy} shows that, in agreement with existing literature, the entanglement entropy of the ground state scales logarithmically with the system size, $S = \frac{1}{3} \ln L + C$, signifying the conformal symmetry with central charge $c = 1$. This property is true for all $| h | < 2$, with only the additive constant $C$ being nonuniversal and depending on $h$.

\begin{figure}[tb]
  \includegraphics[width=1\columnwidth]{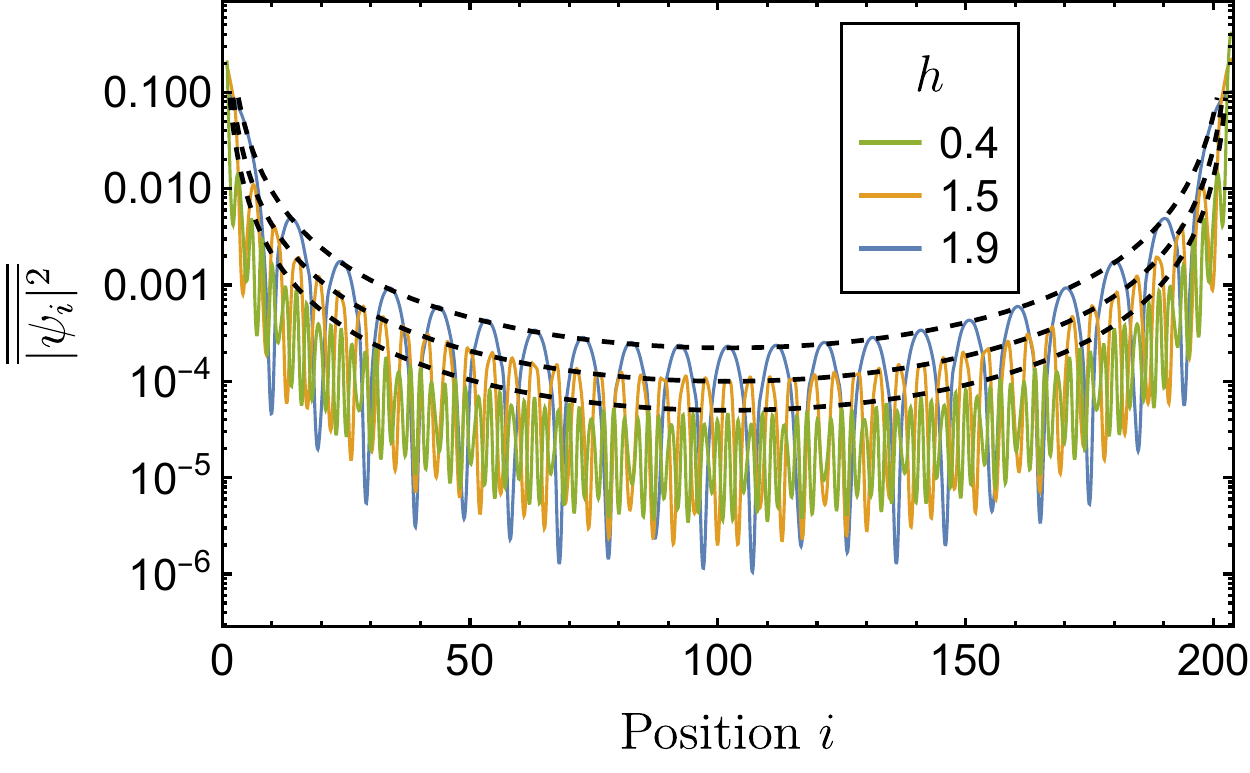}
  \caption{Unscrambled single-particle wave functions for the ground state of the XX model, system size $L = 204$, and different values of transverse field $h$. The system is in a critical phase with conformal symmetry when $h < 2$. The dashed lines are the modulating behavior $\sim \csc^2 (i \pi / L)$.\label{fig:XXwfn}}
\end{figure}

We now investigate the properties of the unscrambled single-particle wave functions. To obtain $U$ from $D$ one can use singular value decomposition, where the singular values should be either 0 or 1. Fig.~\ref{fig:XXwfn} shows the orbitals for a few values of the transverse field $h$ and a set system size $L = 204$. The number of occupied modes is $N = 89$ for $h = 0.4$, $N = 47$ for $h = 1.5$, and $N = 21$ for $h = 1.9$. We find two interesting features of the orbitals. First, the single-particle wave functions show oscillations with the number of troughs equal to $(N - 1)$, which seems to be tightly related to the specifics of the model itself. We leave for future research the peculiar features of these oscillations, such as why each oscillation has a dome-like structure.

Second, the orbitals on average follow a $\overline{\overline{| \psi_i |^2}} \sim \csc^2 (i \pi / L)$ behavior. This can be related to the correlation functions from Eq.~(\ref{eq:corrXX}), which are modulated by $\sim \csc [(i - j) \pi / L]$. In the infinite-volume limit, this turns into a $\sim 1 / (i - j)$ behavior. This brings us to one of properties of the critical phase often used as its signature: the connected correlation functions, defined as $C (r) = | D_{i, i + r} |^2 = \langle n_i \rangle \langle n_{i + r} \rangle - \langle n_i n_{i + r} \rangle$, should decay algebraically as $\sim 1 / r^2$ in the conformal phase. \changes{Note that although the unscrambled wave functions are critical, the single-particle eigenstates of this model will be delocalized.}

In summary, we find that the unscrambled orbitals in the critical conformal phase follow a $\sim \csc^2 (i \pi / L)$ behavior, which we will later use as a marker of criticality.

\subsection{Power-law localization in the symplectic Hatano-Nelson model}
\label{sec:hatano}

\begin{figure}[tb]
  \includegraphics[width=1\columnwidth]{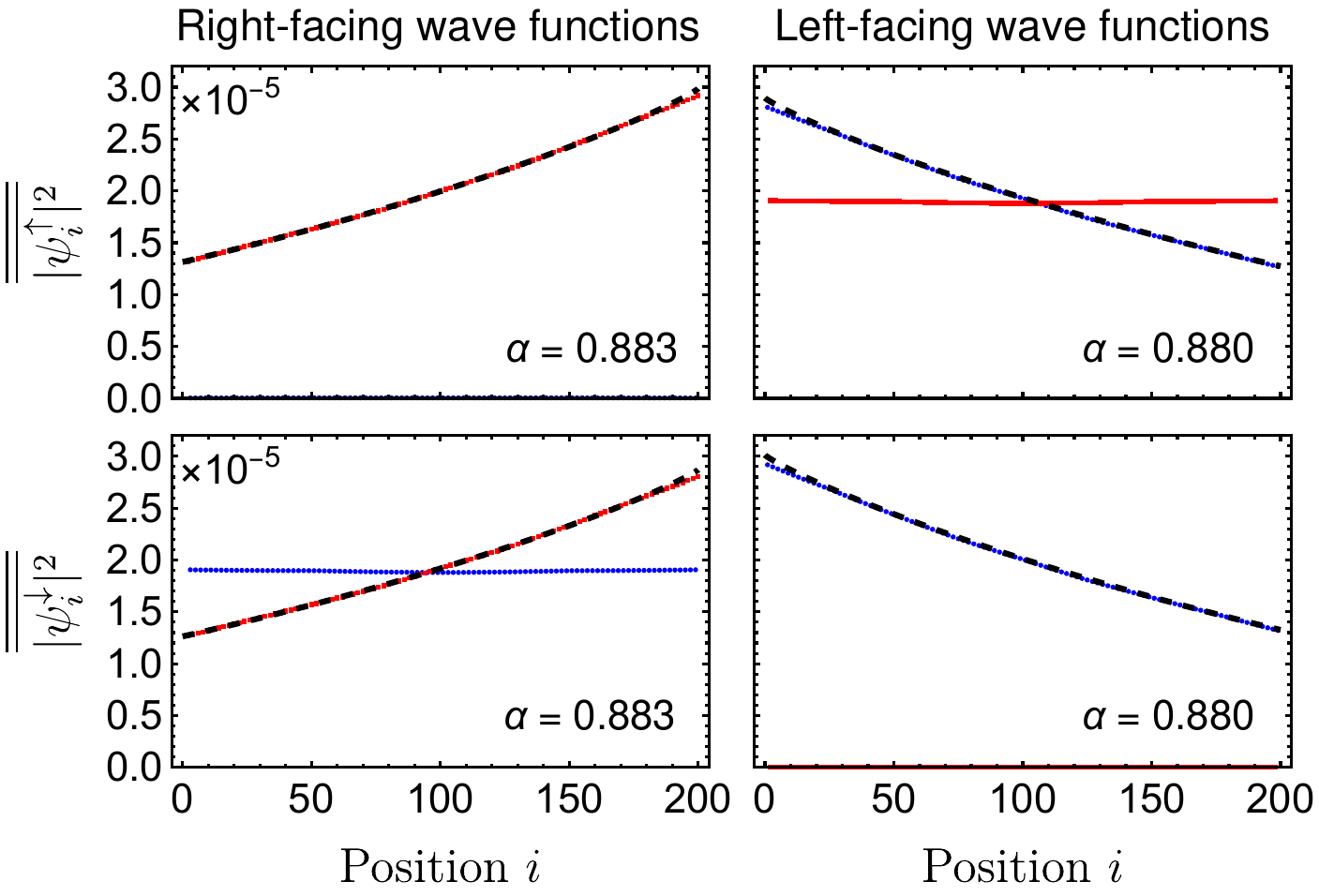}
  \caption{Unscrambled single-particle wave functions of the symplectic Hatano-Nelson model for $L=200$, visibly showing the signatures of the skin effect, with particles either facing right or left. The odd sublattice is indicated in red, while the even sites are denoted in blue. The dashed lines are fits to $\overline{\overline{| \psi_i |^2}} = (a - b | i |^{\alpha})^2$.\label{fig:hatano}}
\end{figure}

To find whether the unscrambling procedure works for a model with power-law localized wave functions, we consider the symplectic Hatano-Nelson model with spinfull particles, previously investigated in Ref.~\cite{Kawabata2023}, given by the Hamiltonian
\begin{align}
  H & = - \frac{1}{2} \sum_{i = 1}^L \left[ \vphantom{\frac{}{}} \Psi^{\dag}_{i + 1} (1 + \mu \sigma_z - i \Delta \sigma_x) \Psi_i \right. \label{eq:hatano} \\
  & \quad \left. \vphantom{\frac{}{}} + \Psi^{\dag}_i (1 - \mu \sigma_z + i \Delta \sigma_x) \Psi_{i + 1} \right], \nonumber
\end{align}
where $\Psi_i = (c_{i \uparrow}, c_{i \downarrow})^{\text{T}}$, often called the Nambu spinor, includes both spin-up and spin-down fermion operators. The Hamiltonian is non-Hermitian, with the strength of non-Hermiticity controlled by $\mu$. Parameter $\Delta$ describes the coupling between spin-up and spin-down modes. The model exhibits a skin effect phase transition~\cite{Kawabata2023}, where the skin effect is present for $| \mu | > | \Delta |$, and absent for $| \mu | < | \Delta |$. At the critical line $| \mu | = | \Delta |$, the states exhibit a skin effect with power-law localized wave functions, which is why we will focus on this case.

\begin{figure*}[t]
  \includegraphics[width=1\textwidth]{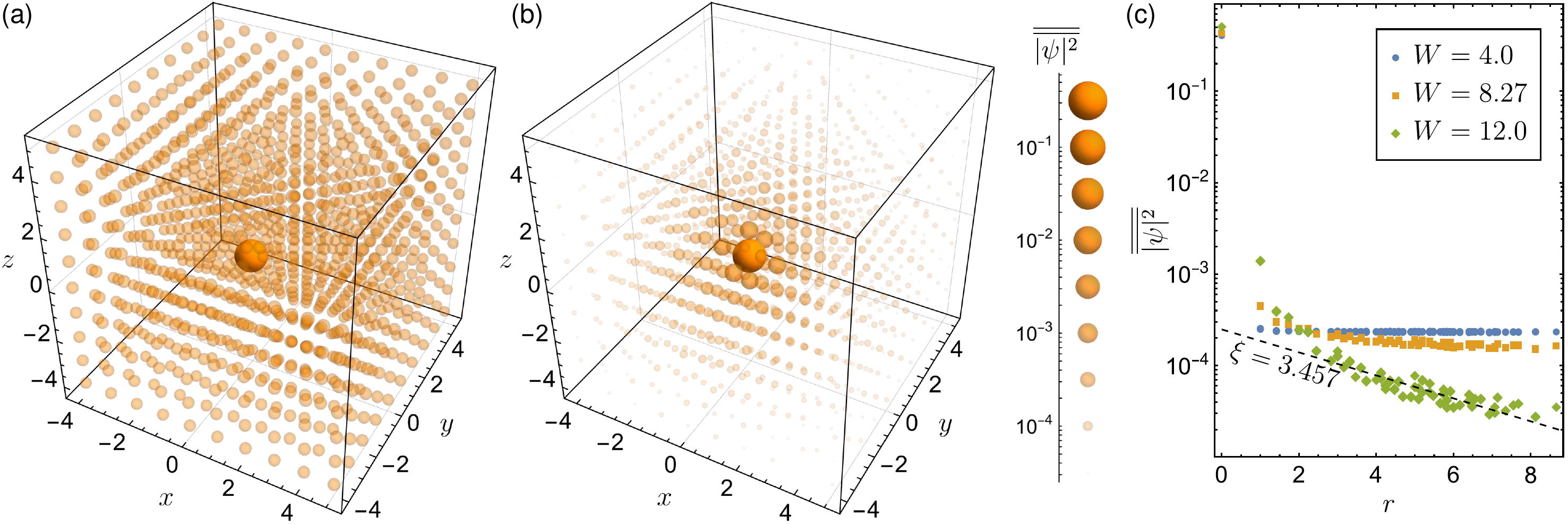}
  \caption{\changes{Unscrambled single-particle wave functions of the 3D Anderson model, with the disorder strength of (a)~$W=4$ (diffusive) and (b)~$W=12$ (localized). We use a lattice of $10 \times 10 \times 10$ sites. The legend in panel (b) applies in panel (a). (c)~Unscrambled orbital dependence on the distance $r = \sqrt{x^2 + y^2 + z^2}$ from the localization center. The dashed line shows the exponential decay $\overline{\overline{| \psi_i |^2}} \sim \exp(-r/\xi)$, where $\xi=3.457$~\cite{MacKinnon1983} is the localization length of the infinite system for $W=12$.} \label{fig:3d_anderson}}
\end{figure*}

The state can be described by a $(2 L \times N)$ matrix $U$, where each row corresponds to spatial-spin index $(i, s), s \in \{ \uparrow, \downarrow \}$. We initialize the state with $N = L = 200$ particles, occupying every site, $i \in \{ 1, \ldots, L \}$, and alternating between spin up and down. We then evolve the system using Eq.~(\ref{eq:tev}), where $\mathbb{H}$ is now a $(2 L \times 2 L)$ matrix encoding the Hamiltonian (\ref{eq:hatano}). We choose the parameters to be at the critical line, $\mu = \Delta = 1$, so that the power-law localization and skin effect are present. After each time step, the wave function needs to be normalized, which can be done by thin QR decomposition of $U=QR$ and setting the new normalized $U' = Q$. We stop the evolution at $t_{\text{fin}} = 100$ and unscramble the orbitals.

We plot the spin-up and spin-down parts of the unscrambled orbitals in Fig.~\ref{fig:hatano}. Each wave function either faces right or left, i.e.\@ its magnitude grows towards the right or towards the left of the system. We can also see that there is a difference between odd and even sites of the chain: one sublattice faces right/left, while the other is either close to zero or constant. This shows an imbalance between the right/left-facing modes, which is the origin of the skin effect. We also show power-law localization by fitting an ansatz $\overline{\overline{| \psi_i |}} \sim a - b | i |^{\alpha}$, where $a, b, \alpha$ are fitting parameters, and $\alpha$ is the power-law exponent. The fitting is done only to the sublattice that is not constant. The fitted power law exponent is consistently found to be $\alpha \approx 0.88$ for all wave functions.

In summary, the unscrambling method yields several significant insights into the underlying properties of the wave functions of the symplectic Hatano-Nelson model. First, unscrambling reveals the existence of two types of quasiparticles -- one traveling right, the other left. Second, it clearly shows the difference between the even and odd sublattices of the system. Third, the unscrambled orbitals are indeed power-law localized, as predicted by the theory.

\changes{
\section{Unscrambling in higher dimensions}
\label{sec:3d-anderson}

One interesting advantage of the unscrambling methodology is that the cost function is essentially independent of the lattice geometry. Consequently, the procedure can be readily extended to higher dimensional lattices. In this section, we apply the unscrambling method to the three-dimensional (3D) Anderson model~\cite{Kramer1993, Bellissard2004, Suntajs2021}. In contrast to the one-dimensional case from the previous section, as well as the two-dimensional case, this model exhibits an Anderson phase transition from a diffusive metallic phase to a localized insulating phase.

We formally define the 3D Anderson model by the Hamiltonian from Eq.~(\ref{eq:anderson}), with the hopping term applied between each link on a cubic lattice of size $L \times L \times L$, and with periodic boundary conditions. The Anderson transition happens at the critical disorder strength of $W_c \approx 8.27$~\cite{Slevin2018}. We initialize the wave function to be half-filled in a 3D checkerboard pattern and evolve the system for time $t_\text{fin} = 8 L$. Subsequently, we perform the unscrambling.

The unscrambled orbitals are shown in Fig.~\ref{fig:3d_anderson}. The averaging is done similarly to the one-dimensional case, where we center each orbital at position $(0,0,0)$ before averaging. The unscrambled single-particle wave functions show a clear change between the diffusive regime [$W = 4$, panel~(a)], where the orbitals are extended across the entire lattice, and the localized phase [$W = 12$, panel~(b)], where the orbitals are localized. In Fig.~\ref{fig:3d_anderson}(c), we show the orbitals as a function of the distance from the localization center, $r = \sqrt{x^2 + y^2 + z^2}$. In the diffusive phase, the orbital is essentially constant, except for $r = 0$, which is an artifact of the averaging procedure. On the other hand, in the localized phase, the orbital exhibits exponential decay $\overline{\overline{| \psi_i |^2}} \sim \exp(-r/\xi)$, where the localization length $\xi = 3.457$ for $W = 12$ is taken from the literature~\cite{MacKinnon1983}. The fit works best away from the localization center, where the behavior is expected to be model-dependent, and away from the maximal distances, where finite-size effects are expected to dominate.

Summarizing, the unscrambling methodology is shown to work in the three-dimensional Anderson model, correctly identifying the presence of the diffusive and localized phases, and predicting the wave function decay consistent with the literature.
}

\section{Single-particle Wave functions in disordered monitored free fermions}
\label{sec:monitored}

We proceed to test the unscrambling methodology in a more complicated scenario: a model of monitored free fermions in a disordered field. This and similar monitored models have been studied extensively in the recent literature~\cite{Cao2019, Chen2020, Alberton2021, Buchhold2021, Jian2020tn, Zhang2021, Tang2021, Botzung2021, Turkeshi2021, Piccitto2022, Turkeshi2022, Tirrito2023, Kells2021, Muller2022, Turkeshi2022negativity, Minato2022, Coppola2022, Popperl2022, Szyniszewski2023, Turkeshi2023, Jian2023, Poboiko2023, Poboiko2023twodim, Chahine2023} in the context of measurement-induced entanglement transitions. Long-time dynamics of the disordered model was found to exhibit an entanglement transition between a conformal critical phase, where the entanglement grows logarithmically with the system size, and an area law phase, where the entanglement saturates to a constant value~\cite{Szyniszewski2023}. Specifically, the critical phase exists for small measurement strengths and small disorder strengths~\footnote{Note that Ref.~\cite{Poboiko2023} suggests that the critical phase in a clean model is a finite-size effect that only exists up to a certain (exponentially large) system size, and beyond this length scale the system is always in an area-law phase. Regardless of whether the theory of Ref.~\cite{Poboiko2023} is applicable to this model, in this work, we never reach this system size, hence we expect to see the signatures of the conformal phase. Also, the effects of the disorder on the theory of Ref.~\cite{Poboiko2023} are currently unknown, especially since the disorder seems to stabilize the critical behavior~\cite{Szyniszewski2023}.}.

In detail, the dynamics of disordered monitored free fermions is governed by the stochastic Schr\"odinger equation,
\begin{align}
  d|\psi(t)\rangle &= -i\, H\, dt|\psi(t)\rangle
  - \frac{\gamma\, dt}{2} \sum_i (n_i-\langle n_i\rangle)^2 |\psi(t)\rangle \nonumber\\
  &\quad +\sum_i (n_i-\langle n_i\rangle) d\eta_i^t|\psi(t)\rangle,
  \label{eq:stoch}
\end{align}
where $\gamma$ is the measurement strength, $d\eta_i^t$ is an It\^o increment (a random number with variance ${\gamma\,dt}$), and the Hamiltonian $H$ is that of the Anderson model from Eq.~(\ref{eq:anderson}). This stochastic evolution models a continuous monitoring of particle number occupation, which can be imagined as a series of homodyne detectors in an experiment~\cite{Alberton2021}. The dynamics can be translated into the following change in the matrix $U$:
\begin{equation}
  U (t + d t) = \mathcal{N} e^\mathbb{M} e^{- i\,dt\,\mathbb{H}} U (t),
  \label{eq:stoch-U-U1}
\end{equation}
where $\mathbb{M}$ is an $L\times L$ measurement matrix with elements $\mathbb{M}_{i j} = \delta_{i j} (d\eta_i^t + (2 \langle n_i \rangle -1) \gamma d t)$, and $\mathcal{N}$ is the normalization constant needed after the measurement (normalization is implemented as in Sec.~\ref{sec:hatano}).

\begin{figure}[tb]
  \includegraphics[width=1\columnwidth]{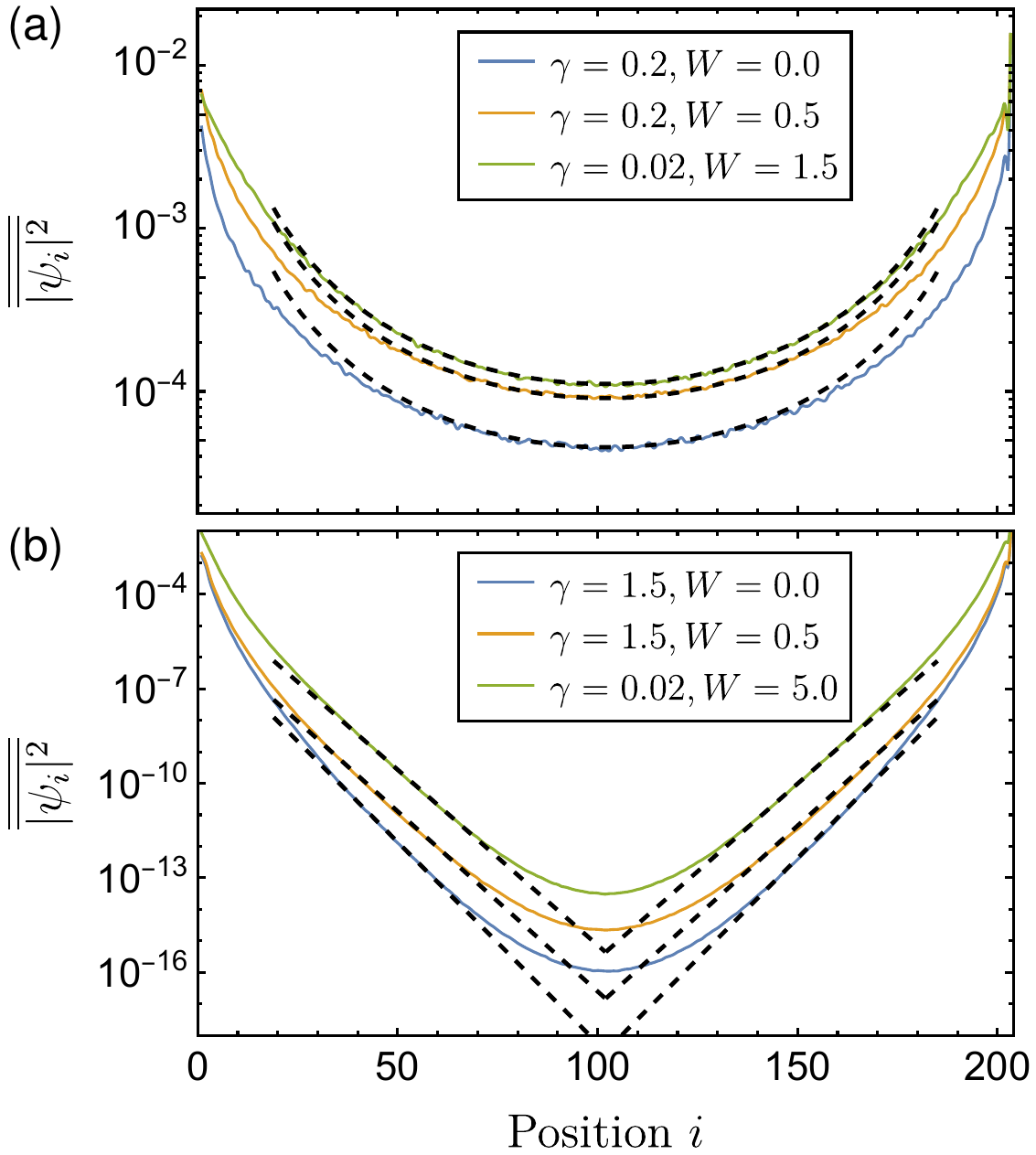}
  \caption{Unscrambled single-particle wave functions of disordered monitored free fermions for the system size of $L=204$. (a)~The parameters used are within the conformally-symmetric phase. The dashed lines show the behavior $\sim \csc^2 (i \pi / L)$. (b)~The parameters used are within the conformally-symmetric phase. The dashed lines show the behavior $\sim \csc^2 (i \pi / L)$.\label{fig:monitored_XX}}
\end{figure}

We initialize the \changes{system} as a half-filled N\'eel state and stop the evolution after the long-time steady state is reached, $t_\text{fin} = 8 L$ for $W\le 1$, and $t_\text{fin} = 8 L W$ when $W > 1$. We subsequently unscramble the resulting wave functions, and plot them in Fig.~\ref{fig:monitored_XX}. Examples of wave functions with parameters chosen inside the critical phase are in Fig.~\ref{fig:monitored_XX}(a), and all fit well within an ansatz $\overline{\overline{| \psi_i |^2}} \sim \csc^2 (i \pi / L)$ (dashed lines). This shows that the wave functions exhibit the generic properties of criticality that we discussed in Sec.~\ref{sec:xx}. The fit does not work well only near the edges of the plot, which correspond to short distances from the center of the wave function, where we indeed do not expect generic behavior. Furthermore, Fig.~\ref{fig:monitored_XX}(b) shows parameters within the area-law phase, where the single-particle wave functions exhibit exponential localization, as in Sec.~\ref{sec:anderson}. Fitting an exponential decay (dashed lines) fails near the (localization) center, but also near the very middle of the plot, where large distances are impacted by finite-size effects related to periodic boundary conditions. A similar behavior was seen in Sec.~\ref{sec:anderson}.

Interestingly, the localized single-particle wave functions show almost no difference in behavior between two regimes: a disorder-dominant regime (where $W$ is large), and a measurement-dominant regime (where $\gamma$ is large instead). Such a difference was hinted at by Ref.~\cite{Szyniszewski2023}, where the orbitals in the disorder-dominant regime appeared to follow power-law localization. However, once unscrambled, the power-law-like behavior transforms into an exponential decay. This accentuates the importance of proper unscrambling in unraveling the underlying properties of the quantum state. Despite this, note that the two regimes remain discernible through distinct decay behaviors of the autocorrelation functions~\cite{Szyniszewski2023}. This echoes similar observations in the interacting case~\cite{Yamamoto2023}.

In conclusion, the unscrambling process provides crucial insights. The unscrambled single-particle wave functions effectively distinguish between conformal and localized phases arising from the interplay of continuous measurements and disorder. This in turn suggests the existence of a measurement-induced transition within the phase diagram of this model. By unscrambling, we gain access to the expected properties of the wave functions, shedding light on the underlying physics.

While the unscrambling methodology has been shown to effectively illuminate the intrinsic properties of single-particle orbitals, it is essential to recognize that all the models considered thus far shared a common feature: particle number conservation. We address this issue in the next section.

\section{Unscrambling without particle number conservation}
\label{sec:z2}

In models lacking particle number conservation, such as those coupled to a superconductor~\cite{Schrieffer1999, DeGennes1999, Mbeng2020}, free-fermionic systems can still be fully described by their two-point correlation functions. In this context, we require particle-conserving elements $D_{i j} = \langle c_i^{\dag} c_j \rangle$, and particle-non-conserving elements $F_{i j} = \langle c_i c_j \rangle$ to comprehensively define these correlations. Moreover, within this framework, there exists an ansatz for any Gaussian state~\cite{Mbeng2020},
\begin{equation}
  | \psi \rangle = \mathcal{N} \exp \left( - \frac{1}{2} \sum_{i j} \left((U^\dag)^{-1} V^{\dag}\right)_{i j} c^{\dag}_i c^{\dag}_j \right) | 0 \rangle,
  \label{eq:z2ansatz}
\end{equation}
where $\mathcal{N} = {\sqrt{| \det U |}}$ is the normalization constant. Matrices $U$ and $V$, both of size $(L \times L)$, describe a Bogoliubov rotation necessary for transforming between standard fermion operators $c_i^\dag$ and Bogoliubov fermions,
\begin{equation}
  \gamma_n^\dag = \sum_i ( U_{i n} c_i^\dag + V_{i n} c_i ).
\end{equation}
Notably, Bogoliubov fermions are quasiparticles composed of particles and holes. The fermion commutation relations impose certain properties on $U$ and $V$. Also, correlation matrices can be expressed as $D = U U^\dag$ and $F = U V^\dag$.

Interestingly, one can easily see that the wave function ansatz in Eq.~(\ref{eq:z2ansatz}) and correlation matrices $D$ and $F$ are invariant under the transformation
\begin{equation}
  U \mapsto U Q, \qquad V \mapsto V Q,
  \label{eq:z2transform}
\end{equation}
where $Q$ is any unitary matrix. This is similar to the freedom in transforming the matrix $U$ in the particle-conserving problem of Eq.~(\ref{eq:u1ansatz}). One can form a matrix
\begin{equation}
  \mathbb{U}= \left(\begin{array}{c}
    U\\
    V
  \end{array}\right),
\end{equation}
of size $2L\times L$, where each column defines the coefficients of a Bogoliubov fermion. Similarly to the particle-conserving problem, we can define the inverse participation ratio of \changes{an orbital} $n$ as
\begin{equation}
  \text{IPR} (n) = \sum_i | \mathbb{U}_{i n} |^4 = \sum_i ( | U_{i n} |^4 + | V_{i n} |^4 ).
\end{equation}
IPR is very clearly representation-dependent, and in general changes when the transformation from Eq.~(\ref{eq:z2transform}) is applied. This is usually not an issue, as $U$ and $V$ often appear directly after the diagonalization of the Hamiltonian in its Bogoliubov-de-Gennes form, i.e.\@ the columns of $U$ and $V$ correspond to its eigenvectors and are well defined. However, if one considers an evolution of the wave function from Eq.~(\ref{eq:z2ansatz}), $U$ and $V$ may no longer be uniquely defined. A paradigmatic example of this is taking a quantum measurement, where both $U$ and $V$ need to be renormalized, losing uniqueness~\cite{Turkeshi2021}.

We can again tackle this issue by unscrambling the Bogoliubov quasiparticles. We propose the following cost function directly associated with the definition of IPR~\footnote{In certain literature (see e.g.\@ Ref.~\cite{Mbeng2020}), the IPR is instead defined as $\text{IPR}(n) = \sum_i (|U_{i n}|^2 + |V_{i n}|^2)^2$. This would lead to an alternative definition of the cost function $f_{n,m} = \sum_i (|U_{i n}|^2 + |V_{i n}|^2) (|U_{i m}|^2 + |V_{i m}|^2)$. The physical meaning of minimizing such a cost function would be to minimize overlaps between sums of particle and hole parts of Bogoliubov fermions. We believe this would not change the resulting wave functions much, and most likely would maintain the localization properties shown for the definition considered in the main text.},
\begin{align}
    f_{n,m} &= \sum_i |\mathbb{U}_{i n}|^2 |\mathbb{U}_{i m}|^2\label{eq:cost_function_xy}\\
    &= \sum_i ( |U_{i n}|^2 |U_{i m}|^2 + |V_{i n}|^2 |V_{i m}|^2),\nonumber
\end{align}
where again minimizing $f = \sum_{n<m} f_{n,m}$ is equivalent to maximizing average IPR. Next, we would like to test the proposed cost function in a realistic model.

\subsection{Disordered monitored free fermions without particle number conservation}

We choose the model of monitored fermions without particle number conservation similar to those in Refs.~\cite{Turkeshi2021, Piccitto2022, Turkeshi2022, Tirrito2023} but with an additional disordered field. Similarly to Sec.~\ref{sec:monitored}, this allows us to test the viability of the method not only in monitored quantum circuits, but also in the presence of the disorder. Comments parallel to Sec.~\ref{sec:monitored} will be applicable here -- we expect that for low values of both measurement strength and the disorder strength, the model will show a critical entanglement behavior, while in other parts of the phase diagram, the area law will be present. The model in question is governed by the modified Anderson Hamiltonian,
\begin{equation}
  H = \sum_{i = 1}^L \left( c_i^{\dag} c_{i + 1} + \kappa c_i^{\dag} c_{i + 1}^{\dag} + \text{h.c.} \right) + \sum_{i = 1}^L h_i n_i,
  \label{eq:anderson-z2}
\end{equation}
where $\kappa$ is the strength of the particle number violating term (the anisotropy in the spin language). The disordered field $h_i$ is again chosen uniformly on $h_i\in [-W, W]$. This Hamiltonian breaks the $U(1)$ symmetry of particle number conservation and instead is endowed with the $\mathbb{Z}_2$ symmetry of fermion parity.

The stochastic Schr\"odinger equation from Eq.~(\ref{eq:stoch}) is now equivalent to the following evolution:
\begin{equation}
  \mathbb{U} (t + d t) = \mathcal{N} e^{2\mathbb{M}} e^{- 2 i\,dt\,\mathbb{H}} \mathbb{U} (t),
  \label{eq:stoch-U-z2}
\end{equation}
where the elements of $\mathbb{H}$ are defined through $H = \Psi^\dag \mathbb{H} \Psi$ with the Nambu spinor $\Psi = (c_1,...,c_L,c^\dag_1,...,c^\dag_L)^T$; similarly for the elements of $\mathbb M$. Note the appearance of the factors of 2 in comparison to Eq.~(\ref{eq:stoch-U-U1})~\cite{Mbeng2020}. We initialize the \changes{system} in the vacuum state $|0\rangle$, which is equivalent to $U = \mathbb{1}$ and $V = \mathbb{0}$, and evolve the state until the steady state is reached, with $t_\text{fin} = 16 L$ when $W \le 1$ and $t_\text{fin} = 16 L W$ for $W > 1$. The anisotropy is set to $\kappa=0.7$. The matrix $\mathbb{U}$ is subsequently unscrambled using the cost function of Eq.~(\ref{eq:cost_function_xy}).

\begin{figure}[tb]
  \includegraphics[width=1\columnwidth]{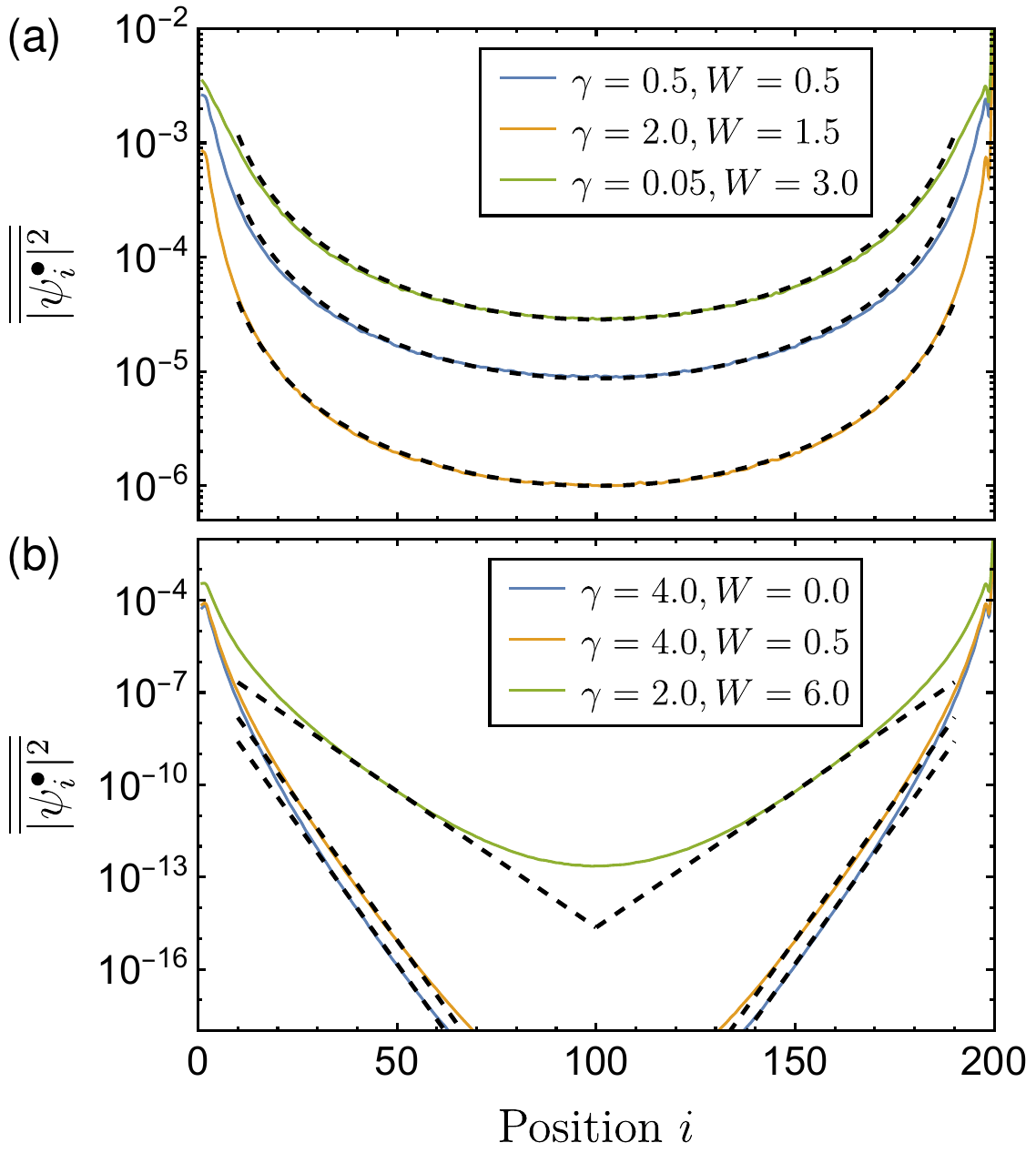}
  \caption{Unscrambled single-particle wave functions of disordered monitored free fermions without particle number conservation for the system size of $L=200$. (a)~The parameters used are within the conformally-symmetric phase. The dashed lines show the behavior $\sim \csc^2 (i \pi / L)$. (b)~The parameters used are within the conformally-symmetric phase. The dashed lines show the behavior $\sim \csc^2 (i \pi / L)$.\label{fig:monitored_XY}}
\end{figure}

\begin{figure}[tb]
  \includegraphics[width=0.9\columnwidth]{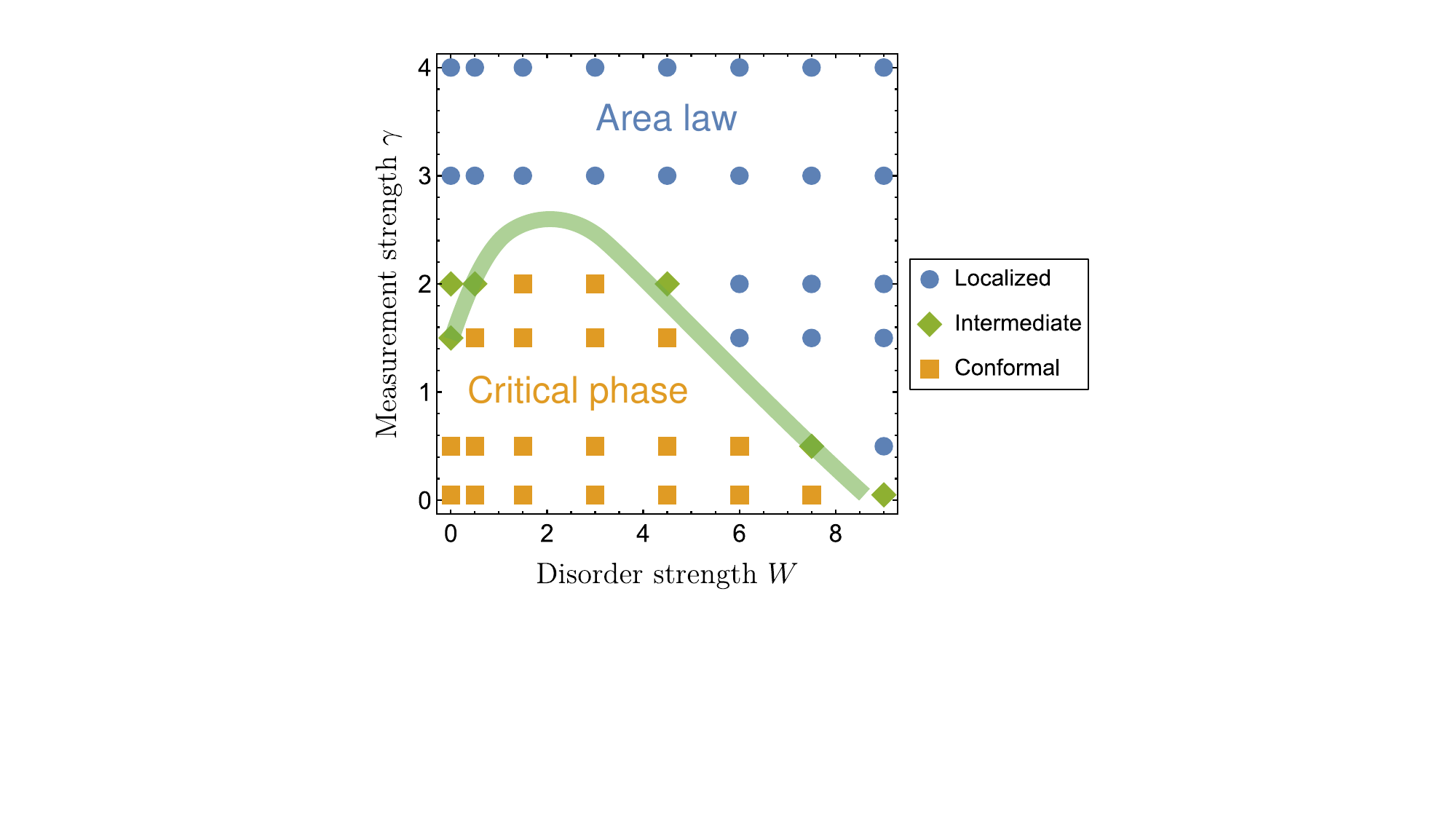}
  \caption{Phase diagram of the disordered monitored free fermions without particle number conservation. The estimate of the transition line (green) is done by estimating whether the behavior of the unscrambled single-particle wave functions is localized (blue circles) or conformal (orange squares). Undecided behavior is designated by green diamonds.\label{fig:monitored_XY_phase_diagram}}
\end{figure}

We present the unscrambled single-particle wave functions for particles $\overline{\overline{|\psi^\bullet_i|^2}}$ (columns of matrix $U$) for $L = 200$ in Fig.~\ref{fig:monitored_XY}, where panel~(a) shows example parameters for which conformal symmetry is present and panel~(b) showcases the area-law phase. The corresponding wave functions for holes $\overline{\overline{|\psi^\circ_i|^2}}$ (columns of matrix $V$) are nearly identical to those of particles, and are therefore not shown in the figures. Surprisingly, the results do not differ significantly from our observations in the particle-conserving case. Specifically, we observe both conformally-symmetric wave functions in Fig.~\ref{fig:monitored_XY}(a) (with clear signatures of $\overline{ \overline{ | \psi^\bullet_i |^2}} \sim \csc^2 (i \pi / L)$ behavior), and exponentially localized orbitals in Fig.~\ref{fig:monitored_XY}(b). Our findings strongly imply that even without particle number conservation, disordered monitored free fermions continue to exhibit a measurement-induced phase transition between an area law and a critical phase with logarithmic entanglement. Indeed, we anticipate that the resulting phase diagram will closely resemble that reported in Ref.~\cite{Szyniszewski2023}. We have approximated how such a phase diagram might appear in Fig.~\ref{fig:monitored_XY_phase_diagram}, where the unscrambled orbitals for $L=200$ were assessed to either fit conformal or localized behavior (see Appendix~\ref{app:phase_diagram_z2} for details). It is important to stress that a rigorous finite-size scaling analysis is necessary to pinpoint the exact location of the transition line. However, this crude estimation emphasizes two important features of this model: the existence of an entanglement transition and the non-monotonicity of the transition line. In previous literature~\cite{Boorman2022, Szyniszewski2023} this non-monotonic behavior has been associated with a mechanism, where a small amount of disorder could facilitate entanglement spreading, stabilizing the critical phase. Our results thus suggest that this mechanism is still present in the model without the $U(1)$ symmetry.

In summary, the unscrambling process effectively extends to free fermions without the conservation of the number of particles. The resulting orbitals serve as powerful indicators of distinct phases and underscore the existence of a measurement-induced transition.

\section{Discussion and outlook}
\label{sec:discussion}

In this work, we have investigated the problem of single-particle representations of free-fermion wave functions. Specifically, our focus was revealing the underlying physics of free-fermion systems with localization. We have developed a method to find a representation in the most localized basis, where the proposed cost function is always decreasing and has a physical interpretation of maximizing the inverse participation ratio of each single-particle orbital. The method works well on paradigmatic examples of 1) conformally symmetric, 2) exponentially localized, and 3) power-law localized systems, and can be extended to encompass systems without particle number conservation. This methodology was shown to be important in understanding the localization transition properties and give insights into the internal structure of free-fermion wave functions. Specifically, the method informs about the properties of the entanglement transition in monitored free-fermion models, and in models with a disorder, where the system exhibits Anderson localization.

Several interesting avenues emerge for further exploration of this work. Firstly, while we have successfully identified clear signatures deep within distinct phases, pinpointing the phase transition line based solely on single-particle wave functions remains a challenge. Therefore, the need for proper finite-size scaling analysis arises. \changes{The scaling could, for example, be performed using the IPR of the unscrambled orbitals, or some other associated property.} It would be fascinating to investigate whether such an analysis can unveil universal properties of the transition, such as critical exponents. Secondly, the method \changes{is} readily extendable to free fermions in two and higher dimensions, where monitoring also forces a phase transition, albeit of a different nature~\cite{Tang2021, Poboiko2023twodim, Chahine2023}.

Thirdly, the question remains whether it is possible to expand the unscrambling methodology to systems beyond non-interacting fermions. Bethe-ansatz solvable integrable models, characterized by states in the Slater determinant form (such as in Ref.~\cite{Carmelo1997}), present an obvious choice for such exploration. Furthermore, while most states cannot be precisely expressed as Slater determinants, many can be approximated as such. It would be intriguing to explore the implications of the unscrambling process for these approximations, and whether this would provide us with any further physical insights. \changes{In the context of many-body quantum systems, the one-particle density matrix (OPDM) description~\cite{Bera2015, Bera2017, Lezama2017, Hopjan2020} has proven useful for understanding many-body localization. Perhaps the unscrambling method could be applied to the natural orbitals (OPDM eigenvectors), although this may necessitate adapting the cost function to incorporate the information about orbital occupations (OPDM eigenvalues). Moreover, the interpretation of this procedure would require further consideration, albeit naively the unscrambling should yield the ``maximally'' localized set of orbitals.} As we venture into the topic of interacting many-body models, the need for \changes{physical} insights becomes increasingly pronounced, together with the need for novel methods that allow us to unravel the intricate dynamics of complex quantum systems.

All relevant data present in this publication can be accessed at~\cite{researchdata}.

\begin{acknowledgments}
  The author would like to thank Christopher J. Turner for useful discussions. M.S.\@ was funded by the European Research Council (ERC) under the European Union's Horizon 2020 research and innovation programme (grant agreement No.\ 853368).
  The author acknowledges the use of the UCL High Performance Computing Facilities (Myriad@UCL and Kathleen@UCL), and associated support services, in the completion of this work.
\end{acknowledgments}

\appendix

\section{Unscrambling method and local minima of the multi-particle cost function}
\label{app:local_min}

In this Appendix, we demonstrate numerically that the unscrambling procedure yields a local minimum of the multi-particle cost function from Eq.~(\ref{eq:cost}). To achieve this, we must apply a transformation that goes beyond the two-particle unscrambling, since we already know that a converged result of the unscrambling process is a simultaneous minimum of all two-particle cost functions. Specifically, consider a unitary matrix,
\begin{equation}
    Q = \exp(i \beta G),
\end{equation}
where $G$ is a Hermitian random matrix from the Gaussian unitary ensemble, and $\beta$ is a parameter controlling the closeness of $Q$ to the identity matrix. Applying $Q$ to the matrix of single-particle wave functions $U$ takes us away from the unscrambling result in a random direction, $U' = U Q$. We then compare the difference $\varepsilon$ between the cost function for $U'$ and that for $U$. If this difference is always positive for any choice of $G$ and $\beta$, it confirms that the unscrambling outcome corresponds to a local minimum.

\begin{figure}[t]
  \includegraphics[width=\columnwidth]{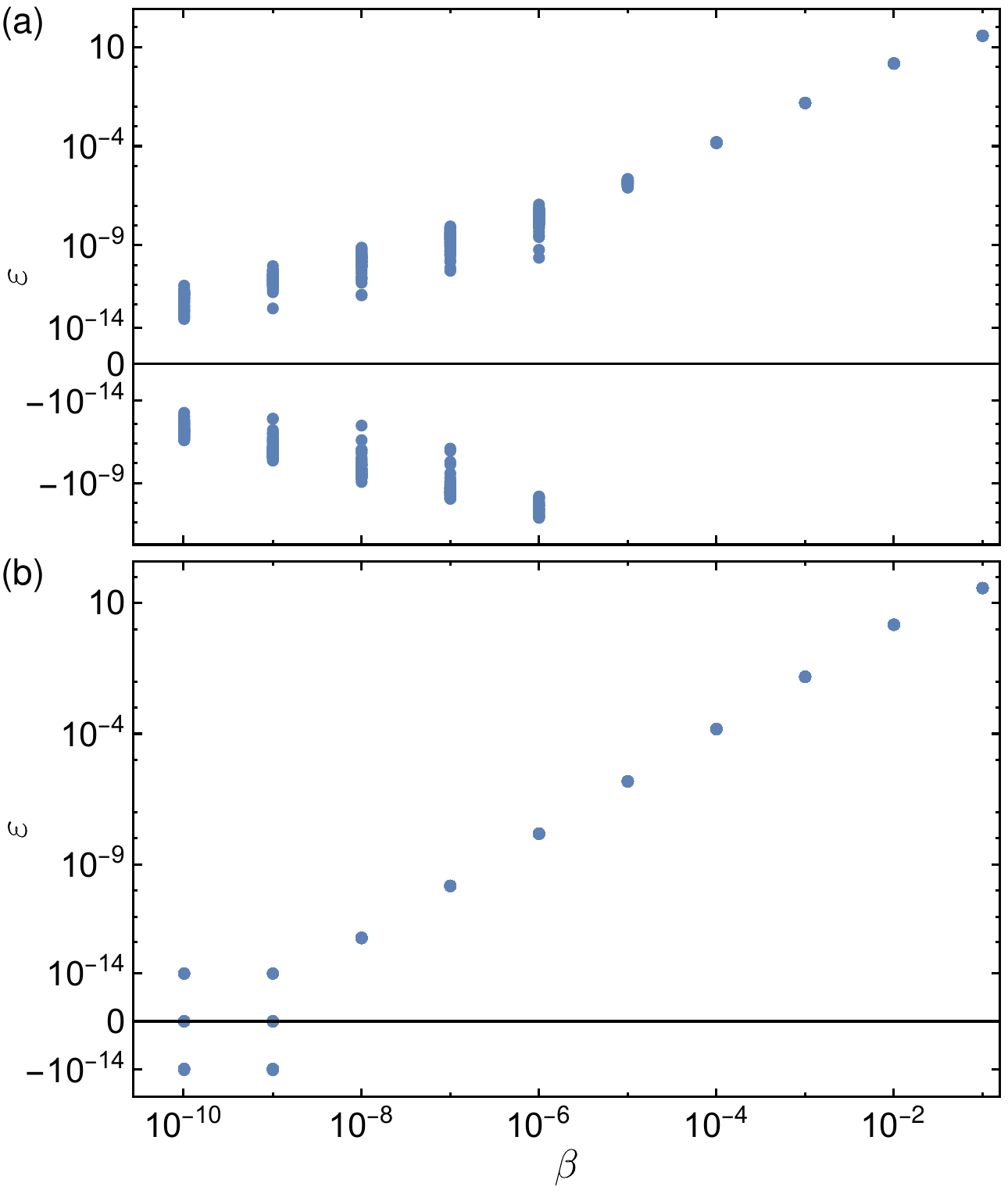}
  \caption{Difference $\varepsilon$ of cost functions between transformed single-particle orbitals $U' = \exp(i \beta G)$ and the original $U$. (a)~Example of the unscrambling when the result has not converged. $\varepsilon$ may be negative for some random matrices $G$ for low enough $\beta$, signifying that the result is not a minimum. (b)~Converged result is numerically a local minimum, with $\varepsilon$ always positive. All panels use results for the disordered monitored free fermions without particle number conservation (see Sec.~\ref{sec:z2}) for $\gamma = 0.05, W = 4.5$, and show 100 random realizations of $G$.  \label{fig:local_minimum}}
\end{figure}

To illustrate how an unscrambling result would fail this test, we stop the unscrambling before the convergence is reached, with the corresponding plot shown in Fig.~\ref{fig:local_minimum}(a). For sufficiently small $\beta$, the values of $\varepsilon$ may be negative, indicating that the unconverged result is not a minimum. In contrast, Fig.~\ref{fig:local_minimum}(b) showcases an example of a converged result, where $\varepsilon$ consistently remains positive (except when the difference is smaller than the numerical precision of $\sim 10^{-14}$), suggesting a local minimum. We have rigorously examined samples from all models considered in this work, and they all fall in the latter category. Thus, we are reasonably confident that the result of the unscrambling process always corresponds to a local minimum of the cost function.

\changes{\section{Single-particle eigenvectors of the Anderson model}
\label{app:eigen_vs_evol}

\begin{figure}[t]
  \includegraphics[width=1.0\columnwidth]{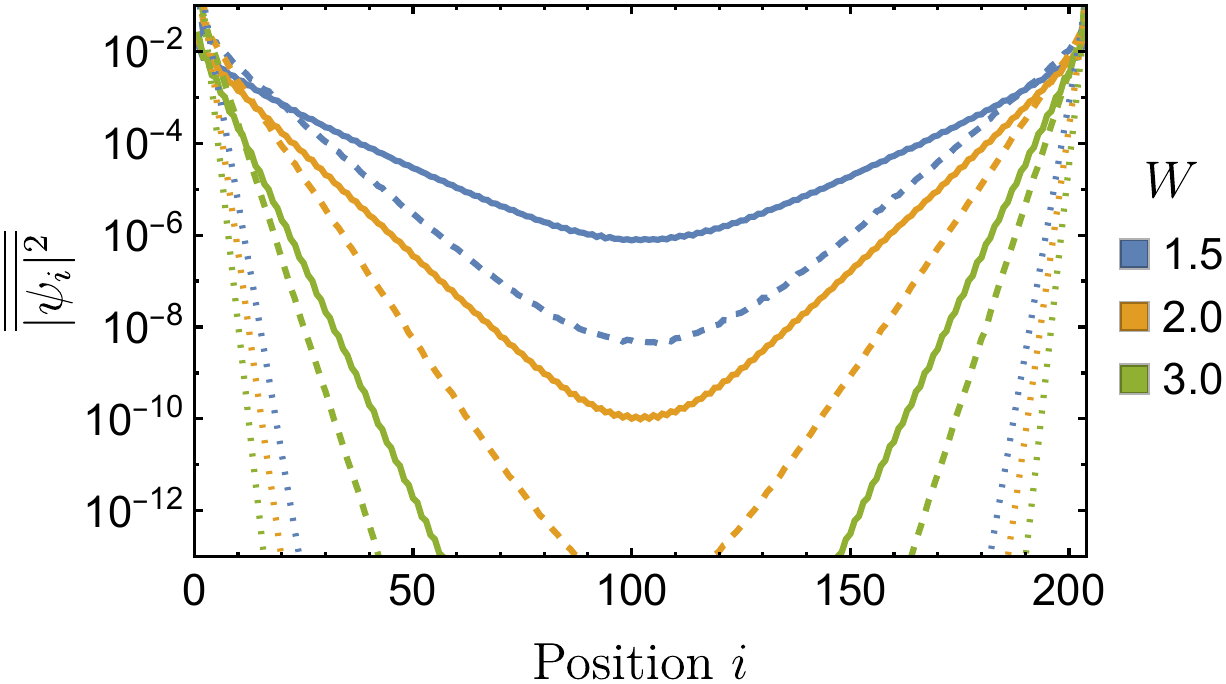}
  \caption{\changes{Unscrambled orbitals (solid lines) of the one-dimensional Anderson model on a periodic chain of size $L=204$, compared with the averaged single-particle eigenfunctions in the middle 10\% (dashed lines) and edge 5\% (dotted lines) of the spectrum.} \label{fig:anderson_comparison}}
\end{figure}

In this Appendix, we compare the behavior of single-particle eigenvectors of the 1D Anderson model with the unscrambled orbitals of Sec.~\ref{sec:anderson}. The eigenvectors can be obtained by diagonalizing the Hamiltonian matrix $\mathbb{H}$ of size $L \times L$, defined below Eq.~(\ref{eq:tev}). Conceptually, the evolution of the orbitals from Sec.~\ref{sec:anderson} explores the bulk of the Hilbert space, hence we expect the behavior to be representative of the system near $E \approx 0$ (middle of the spectrum). However, the single-particle eigenvectors of the Anderson Hamiltonian are quite special points of the Hilbert space, hence it is hard to make a direct comparison. Fig.~\ref{fig:anderson_comparison} shows that the single-particle eigenfunctions generically decay more rapidly than the unscrambled orbitals. Single-particle eigenvectors at the edge of the spectrum exhibit a more rapid decay than those near the middle of the spectrum ($E \approx 0$).}

\section{Phase diagram of the disordered monitored free fermions without particle number conservation}
\label{app:phase_diagram_z2}

\begin{figure}[t]
  \includegraphics[width=0.9\columnwidth]{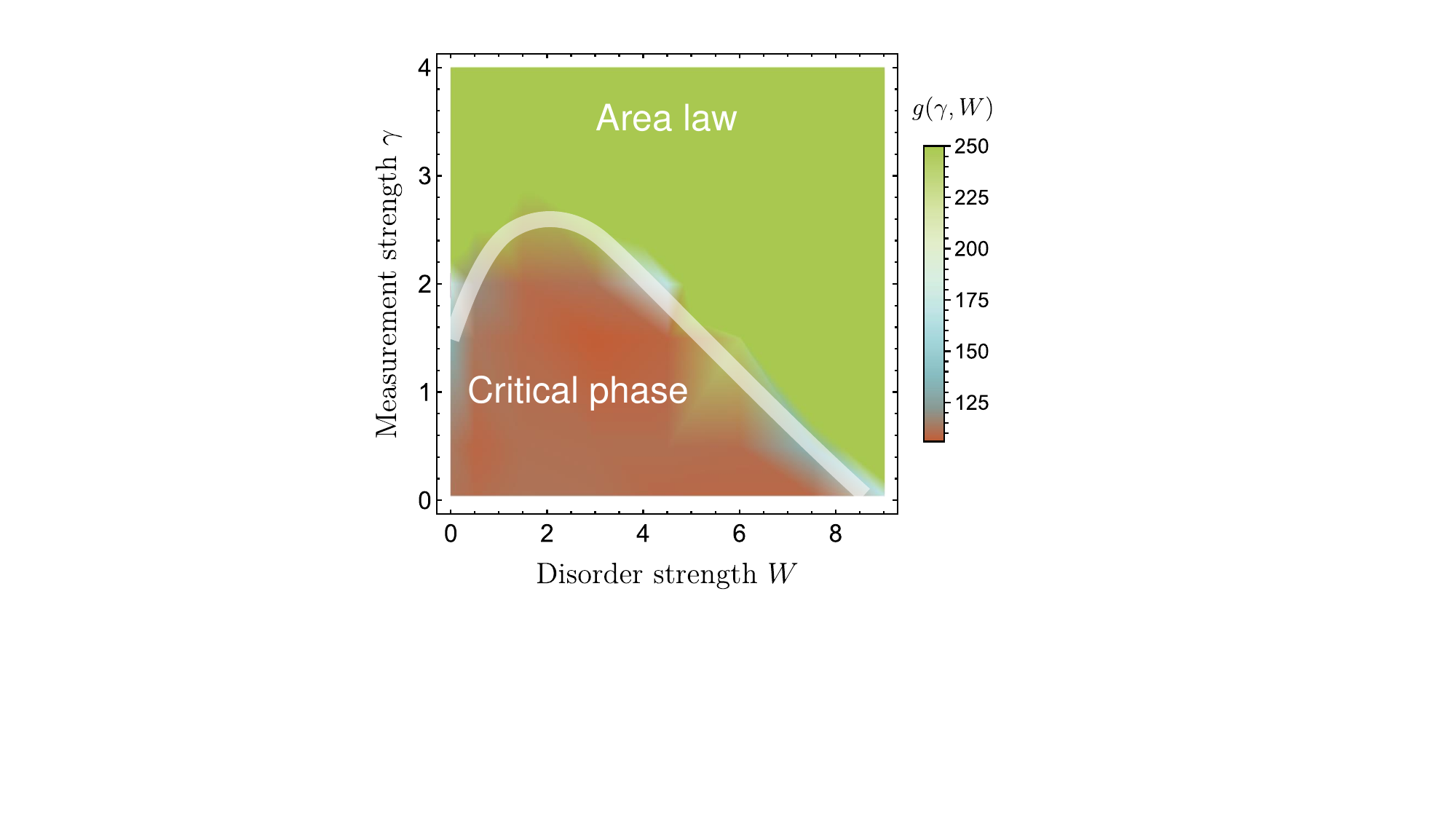}
  \caption{Density plot of the function $g$ that measures the difference between the results and the conformal behavior for the disordered monitored free fermions without particle number conservation. The transition line (in white) estimates the transition point between the conformal behavior and the area law. \label{fig:monitored_XY_phase_diagram_fit}}
\end{figure}

Here we show in detail how the localized, conformal, and intermediate behaviors were determined in Fig.~\ref{fig:monitored_XY_phase_diagram}. In order to decide whether the conformal behavior is present, we fix the coefficient $A$ of the function $\psi = A \csc^2(i \pi/L)$ by choosing it to fit exactly to the data at the point $i=L/4$. The choice is so that we ignore the nonuniversal behavior near the orbital center ($i = 0$) and the finite-size effect near $i=L/2$. Then, a function $g(\gamma, W)$ measures the difference between the data and the expected behavior on a log scale,
\begin{equation}
    g(\gamma, W) = \sum_{i \in \mathcal{L}} \left|
    \ln(\overline{\overline{|\psi^\bullet_i|^2}}) - \ln[A \csc^2(i \pi/L)]
    \right|,
\end{equation}
where $\mathcal L$ encompasses the middle 80\% of the sites. We then perform this calculation for a grid $W \in$ \{0.0, 0.5, 1.5, 3.0, 4.5, 6.0, 7.5, 9.0\}, $\gamma \in$ \{0.05, 0.5, 1.5, 2.0, 3.0, 4.0\} at the system size of $L = 200$. We do not consider the line $\gamma = 0$ corresponding to an Anderson-localized model, since it most likely follows a separate behavior mimicking Ref.~\cite{Szyniszewski2023}, where there is an abrupt change as soon as any monitoring is introduced.

The result is shown in Fig.~\ref{fig:monitored_XY_phase_diagram_fit}. The points in the red region are chosen as being close to critical, the green region is localized, and the blue region points are designated as intermediate. The transition line is then estimated as the boundary between the red and green regions.

\bibliography{refs}

\end{document}